\documentclass[sigconf]{acmart}

\AtBeginDocument{%
  \providecommand\BibTeX{{%
    \normalfont B\kern-0.5em{\scshape i\kern-0.25em b}\kern-0.8em\TeX}}}

\setcopyright{acmcopyright}
\copyrightyear{2021} 
\acmYear{2021} 
\setcopyright{acmlicensed}\acmConference[CODS COMAD 2021]{8th ACM IKDD CODS and 26th COMAD}{January 2--4, 2021}{Bangalore, India}
\acmBooktitle{8th ACM IKDD CODS and 26th COMAD (CODS COMAD 2021), January 2--4, 2021, Bangalore, India}
\acmPrice{15.00}
\acmDOI{10.1145/3430984.3431966}
\acmISBN{978-1-4503-8817-7/21/01}

\begin{document}

\title{Trustworthy AI}


\author{Richa Singh}
\affiliation{%
  \institution{IIT Jodhpur, India}}  
  \email{richa@iitj.ac.in}

\author{Mayank Vatsa}
\affiliation{%
  \institution{IIT Jodhpur, India}}
\email{mvatsa@iitj.ac.in}

\author{Nalini Ratha}
\affiliation{%
  \institution{University of Buffalo, USA}}
\email{nratha@buffalo.edu}

\renewcommand{\shortauthors}{Singh, et al.}

\begin{abstract}

  Modern AI systems are reaping the advantage of novel learning methods. With their increasing usage, we are realizing the limitations and shortfalls of these systems. Brittleness to minor adversarial changes in the input data, ability to explain the decisions, address the bias in their training data, high opacity in terms of revealing the lineage of the system, how they were trained and tested, and under which parameters and conditions they can reliably guarantee a certain level of performance, are some of the most prominent limitations. Ensuring the privacy and security of the data, assigning appropriate credits to data sources, and delivering decent outputs are also required features of an AI system.
  We propose the tutorial on “Trustworthy AI” to address six critical issues in enhancing user and public trust in AI systems, namely: (i) bias and fairness, (ii) explainability, (iii) robust mitigation of adversarial attacks, (iv) improved privacy and security in model building, (v) being decent, and (vi) model attribution, including the right level of credit assignment to the data sources, model architectures, and transparency in lineage.

%Modern AI systems while reaping the advantage of novel learning methods, they exhibit brittleness to minor changes in the input data and lack the capability to explain their decisions to a human. Furthermore, they are unable to address the bias in their training data, as demonstrated in non-uniform performance across different groups, and are often highly opaque in terms of revealing the lineage of the system. Present AI systems have not demonstrated the ability to learn without compromising on the privacy and security of data or being decent. Nor can they even assign appropriate credit to the data sources. The tutorial on Trustworthy AI addresses these critical issues in enhancing user and public trust in AI systems. 
\end{abstract}





\begin{teaserfigure}
  \includegraphics[width=1.05\textwidth]{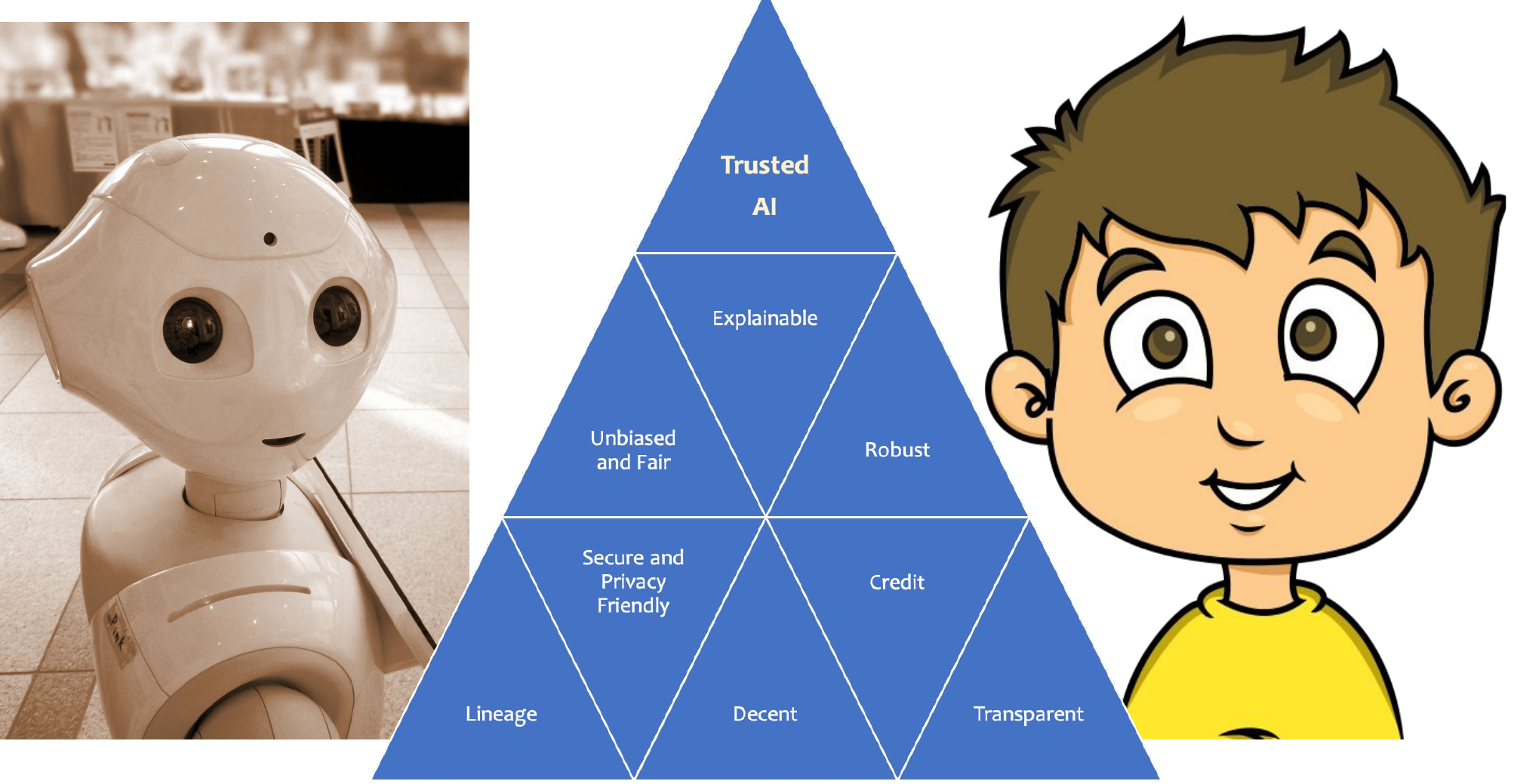}
  \caption{Components of AI Systems being Trustworthy and Decent.}
  \label{fig:teaser}
\end{teaserfigure}

\maketitle

\section{Introduction}

%In every walk of life, AI is playing a significant and increasing role. AI systems are being employed for making mundane day to day decisions such as healthy food choices and dress recommendation, as well as mission critical and life-changing decisions such as diagnosis of diseases, detection of financial frauds, and selecting new employees. Many upcoming applications such as autonomous driving, automated financial loan approval and cancer treatment recommendations have many worrying about the level of trust associated with AI today. Such concerns are genuine as many weaker sides of modern AI systems have been exposed through adversarial attacks, bias, and lack of explainability in the current rapidly evolving AI systems. 

Artificial intelligence (AI) systems are becoming an integral part of our daily lives. They are being employed for making mundane day to day decisions such as healthy food choices and dress recommendation, as well as for important and impactful decisions such as diagnosis of diseases, detection of financial frauds, and selecting new employees. Their increasing deployment in upcoming applications such as autonomous driving, automated financial loan approval, and cancer treatment recommendations have many worrying about the level of trust associated with AI today. Such concerns are genuine as many weaker sides of modern AI systems have been exposed through adversarial attacks, bias, and lack of explainability in the current rapidly evolving AI systems. Therefore, it is important to build mechanisms and approaches for ``Trustworthy AI'' systems. 

Building a trustworthy AI system requires understanding if the model is biased or not. Bias has been a critical Achilles’ heel problem for modern AI systems. Many applications from face recognition to language translation have shown high levels of bias in the systems, as demonstrated in non-uniform performance across different groups and test sets. This has strong implications on the fairness and accountability of such systems, with extremely significant societal implications. Explainability and interpretability are a requirement for such systems in many different contexts, for example law enforcement and medical, where black box decision making is not acceptable. Even though modern AI systems have reported high levels of accuracy, they are unable to explain their decision process and the causes of failures or successful cases to humans. Privacy and security are critical to success of AI beyond high accuracies. Recent research has shown that AI algorithms can exploit information extracted from social media to de-anonymize blurred faces, and promote unwanted spying through surveillance cameras. Such AI applications present both challenges and opportunities: while surveillance systems enhance safety of individuals and society at large, their susceptibility to attacks and breaches also provide the opportunity for abuse. Adversarial attacks in particular have created a huge negative perception with users that AI systems can be fooled easily. As researchers, we need to establish and promote a rigorous framework to formulate the problems in adversarial machine learning, evaluate the impact and consequences under various adversarial attacks, and characterize the properties that ensure the security of AI models.

As it has been observed in many areas, openness helps in unlocking larger potentials. Many of the AI systems do not disclose the lineage of the model, training data and performance details. More research is needed to address a common minimum acceptable practice to be disclosed by the systems. Data and model attribution are critical components of trusting an AI system. Accurate descriptions of training data, architecture and reliable test conditions are critical to guarantee a level of performance within predefined ranges, set users expectations and potentially explain potential biases and failures. Furthermore, especially in complex AI systems made of multiple components, attribution of a given prediction or signal from one specific model is fundamental for explainability, as well as security. Being able to reliably recognize the signature of an AI system provides a robust method to identify tampering, errors and potential breaches.

Furthermore, AI systems have barely benefited from recent advances in security such as fully homomorphic encryption. Recently, blockchain technology has been revolutionizing the way transactions are conceived, executed, managed, and monetized. While the commercial benefits of blockchain infrastructure are imminent, the applicability and advantages that the technology can offer to AI researchers and systems is still to be explored. %Of specific interest to researchers and application developers in AI is the tremendous opportunity to make a connection to these emerging infrastructure capabilities and realize how their skills can be leveraged to make an impact by marrying AI and blockchain technologies. 
As the world moves towards increasing decentralization of AI and emergence of AI marketplaces, a blockchain-based infrastructure would be essential to create the necessary trust between diverse stakeholders.

Another important aspect of being a human like AI system is being ``decent''; though it has not being discussed in the literature directly. While designing an AI system, it is expected that the social behaviors that are habitual in human to human interactions are followed in AI agent-human interactions. We generally train on a noisy corpus and do not train/enforce it to be decent, which may lead to indecent responses by the agent. Several instances in the past have shown indecent behavior by the AI systems \cite{chatbot,metoo,hotel} and many of such systems have been decommissioned due to their indecent responses. Therefore, we believe that in order to build a trustworthy AI system, it is very important to have ensure that the system is \textit{decent}. 

Although researchers are studying individual challenges of AI systems independently, we believe that we need a comprehensive attempt to bring together all the aspects of trustworthy in AI under one umbrella and understand the interlinkages between them. This tutorial will discuss the fundamentals of trustworthy and decent AI, and summarize key and recent contributions in this area as well as open challenges. As shown in in Figure \ref{fig:teaser}, we focus on the following topics:

\begin{itemize}
\item Bias and Fairness (estimation and mitigation)
\item Adversarial Robustness (detection and mitigation)
\item Explainability and Interpretability
\item Trustworthiness and Security via Blockchain
\item Privacy Preserving AI
\item Credit, Lineage, and Transparency Approaches
\item Decent AI
\end{itemize}

\section{Building Trusted/Trustworthy AI Systems}

The EU guidelines on defining Trustworthy AI lays out seven requirements \cite{eu-guidelines}:
\begin{enumerate}

\item ``\textit{Human agency and oversight, including fundamental rights, human agency and human oversight}".

\item ``\textit{Technical robustness and safety, including resilience to attack and security, fallback plan and general safety, accuracy, reliability and reproducibility}".

\item ``\textit{Privacy and data governance, including respect for privacy, quality and integrity of data, and access to data}".

\item ``\textit{Transparency, traceability, explainability and communication}".

\item ``\textit{Diversity, non-discrimination and fairness, including the avoidance of unfair bias, accessibility and universal design, and stakeholder participation}".

\item ``\textit{Societal and environmental well-being, sustainability and environmental friendliness, social impact, society and democracy}".

\item ``\textit{Accountability, auditability, minimisation and reporting of negative impact, trade-offs and redress}".
\end{enumerate}

There are several directions that researchers have pursued for addressing the above mentioned components. For instance, bias and fairness research has focused on understanding and estimating bias as well as mitigating and accounting for bias \cite{ntoutsi2020bias, buolamwini2018gender, celis2019learning, nagpal2019deep, 10.3389/fdata.2020.00018,creager2019flexibly, nagpal2020attribute}. It has been discussed in great detail in several papers \cite{mehrabi2019survey,drozdowski2020demographic,ntoutsi2020bias}. Similarly, research on robustness against adversarial attacks has focused on generating attack models, detecting adversarial perturbations, and mitigating these attacks \cite{7958568, goswami2018unravelling,xie2017mitigating,liu2018towards,xu2017feature, agarwal2018robustness,goswami2019unravelling,liu2019detection,lu2017safetynet, 9207872, goodfellow6572explaining, madry2017towards, tramer2017ensemble, shafahi2019adversarial, agarwal2020sign}. Recently, different surveys have also highlighted the research progress in adversarial robustness \cite{8611298, ren2020adversarial, singh2020robustness}. In recent research threads, deep learning models are protected via cryptography measures to convert the internal layers of CNN into blocks of the blockchain \cite{goel2019robustness,goel2019robustness2,goel2020robustness}. Researchers are also working towards providing certified defense against adversarial perturbations. However, most of them are evaluated on the first-order adversary or gray-scale images \cite{wong2018scaling,raghunathan2018semidefinite}. Cohen et al. \cite{cohen2019certified} have shown the certified robustness of large scale ImageNet images and proved slight robustness for $l_2$ attacks. Moreover, research suggests that adversarial training is vulnerable to black-box attacks with several privacy issues and creates blind-spots for further attacks \cite{zhang2019limitations,mejia2019adversarial}. In other research directions, explainability \cite{xai, xai2, lin2020see}, transparency \cite{haibe2020transparency}, data lineage \cite{lineage}, privacy and security \cite{ChhabraSVG18}, and social well being \cite{shi2020artificial} have also been addressed.

\begin{figure*}[]
\begin{center}
   \includegraphics[width=0.8\linewidth]{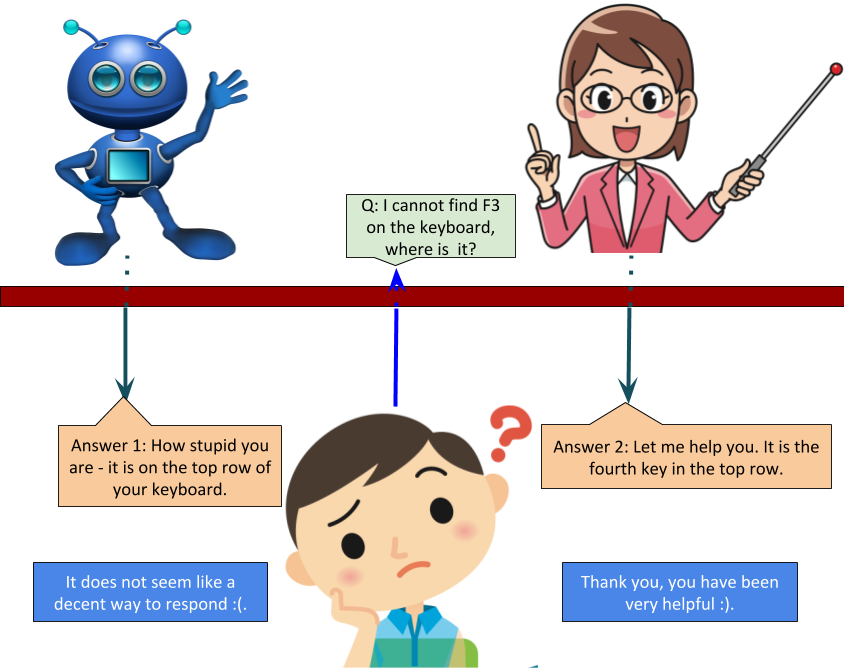}
\end{center}
   \caption{An illustration of ``Decent AI Turing Test.''}
\label{fig:turing}
\end{figure*}

\subsection{Comparison with Similar Concepts}

\begin{itemize}
\item{Robust AI:}
In computer science, robustness is defined as the ``\textit{ability of a computer system to cope with errors during execution and cope with erroneous input}" \cite{wiki-robust}. 
\item{Ethical AI:}
The ethics of artificial intelligence, as defined in \cite{wiki-ethics}, ``\textit{is the part of the ethics of technology specific to robots and other artificially intelligent entities. It can be divided into a concern with the moral behavior of humans as they design, construct, use and treat artificially intelligent beings, and machine ethics, which is concerned with the moral behavior of artificial moral agents (AMAs). It also includes the issues of singularity and superintelligence.}" An AI system that follows this ethics guidelines would be considered as an Ethical AI system.
\item{Fair (unbiased) AI:}
``\textit{In machine learning, a given algorithm is said to be fair, or to have fairness, if its results are independent of the given variables, especially those considered sensitive, such as the traits of individuals which should not correlate with the outcome (i.e. gender, ethnicity, sexual orientation, disability, etc)}" \cite{wiki-fair}. We would call an AI system fair or unbiased if it meets this criterion.
\item{Safe AI:} Safe AI can be  defined as a system that supports AI safety \cite{safe-ai} - ``\textit {Artificial Intelligence (AI) Safety can be broadly defined as the endeavour to ensure that AI is deployed in ways that do not harm humanity}".

\item {Dependable AI:} Dependable AI is defined as a system which is reliable, verifiable, explainable, and secure. This connects above mentioned themes and expands on the reliability and verifiability.   
\end{itemize}

\section{Decent AI}

We postulate that a truly intelligent system should be decent. In this context, we try to explore how can machines be trained to exhibit decent behavior. In order to define how to evaluate an AI system to be decent or indecent, we present the concept of Decent AI Turing Test. 

\noindent \textbf{Decent AI Turing Test:} Inspired from the Turing test, as shown in Figure \ref{fig:turing}, we extend the original Turing test to Decent AI Turing Test. A human evaluator poses a question to the system and s/he does not know if the answer is given by a decent human agent or an automated AI agent. The response is then judged by the evaluator if it is correct or not and if it is decent or not. If the evaluator is not able to reliably differentiate the AI agent's response with the decent human agent, the AI agent is said to have passed the Decent AI Turing Test. More formally, we define: 

\textit{A Decent AI Turing Test is a method of inquiry in AI for determining whether or not a computer is capable of thinking and behaving like a decent human being.}

With this definition, we can argue that Decent AI is conceptually very different and complements all the characteristics of trustworthy AI systems. By that we mean that a decent AI system can have the properties of being fair, dependable, and trusted additionally. 

\section{Research Questions}

There are a series of open research questions that can be explored. For instance: 

\begin{enumerate}
    \item How to automatically measure/evaluate trustworthiness and decency of an AI system?
    \item How to integrate trustworthiness and decency with the performance/accuracy of the system?
    \item How to analyze the relationship between decency and different factors of trustworthy and dependable AI \cite{ToreiniACEZM20}?
    \item How to democratize trustworthy and decent AI systems? 
\end{enumerate}

\section{Brief Biographies} 

\textbf{Richa Singh} received the M.S. and Ph.D. degree in computer science from West Virginia University, Morgantown, USA. She is currently a Professor at IIT Jodhpur, India. She is a Fellow of IAPR and a Senior Member of IEEE and ACM. She was a recipient of the Kusum and Mohandas Pai Faculty Research Fellowship at the IIIT-Delhi, the FAST Award by the Department of Science and Technology, India, and several best paper and best poster awards in international conferences. She is/was the Program Co-Chair of IJCB2020, FG2019 and BTAS 2016, and a General Co-Chair of FG2021 and ISBA 2017. She is also the Vice President (Publications) of the IEEE Biometrics Council and an Associate Editor-in-Chief of Pattern Recognition.

\textbf{Mayank Vatsa} received the M.S. and Ph.D. degrees in Computer Science from West Virginia University, USA. He is currently a Professor with IIT Jodhpur, India, and the Project Director of the Technology and Innovation Hub on Computer Vision and Augmented \& Virtual Reality under the National Mission on Cyber Physical Systems by the Government of India. He is the recipient of the prestigious Swarnajayanti Fellowship from the Government of India, the A. R. Krishnaswamy Faculty Research Fellowship at the IIIT-Delhi, and several best paper and best poster awards at international conferences. He is an Area/Associate Editor of Information Fusion and Pattern Recognition, the General Co-Chair of IJCB 2020, and the PC Co-Chair of IEEE FG2021. He has also served as the Vice President (Publications) of the IEEE Biometrics Council where he started the IEEE Transactions on BIOM.

\textbf{Nalini K. Ratha}  is an Empire Innovation Professor of computer science and engineering with the University at Buffalo (UB), State University of New York, Buffalo, NY, USA. He received his B. Tech. in Electrical Engineering from IIT Kanpur, M.Tech. degree in Computer Science and Engineering also from IIT Kanpur and Ph. D. in Computer Science from Michigan State University. He has authored more than 100 research papers in the area of biometrics and has been co-chair of several leading biometrics conferences and served on the editorial boards of IEEE TPAMI, IEEE TSMC-B, IEEE TIP and Pattern Recognition journal. He has co-authored a popular book on biometrics entitled Guide to Biometrics and also co-edited two books entitled Automatic Fingerprint Recognition Systems and Advances in Biometrics: Sensors, Algorithms and Systems. He has offered tutorials on biometrics technology at leading IEEE conferences and also teaches courses on biometrics and security. He is Fellow of IEEE, Fellow of IAPR and an ACM Distinguished Scientist. During 2011-2012 he was the president of the IEEE Biometrics Council. He was awarded the IEEE Biometrics Council leadership award in 2019. 

\begin{acks}
M. Vatsa is partially supported through the Swarnajayanti Fellowship from the Government of India. R. Singh and M. Vatsa are partially supported by a research grant from the Ministry of Electronics and Information Technology, Govt. of India.
\end{acks}

\bibliographystyle{ACM-Reference-Format}
\bibliography{sample-base}


\end{document}